\documentclass[]{article}
\usepackage{graphicx,amsmath}
\usepackage[hmargin=3cm,vmargin=3cm,bindingoffset=0.5cm]{geometry}
\usepackage{amsmath,amssymb,dsfont}
\usepackage{authblk}

\usepackage{url}
\usepackage{xcolor}

\title{A Discrete View of the Indian Monsoon to Identify Spatial Patterns of Rainfall}
\author[1]{Adway Mitra}
\author[2]{Amit Apte}
\author[2]{Rama Govindarajan}
\author[2]{Vishal Vasan}
\author[3,4]{Sreekar Vadlamani}
\affil[1]{School of Electrical Sciences, Indian Institute of Technology, Bhubaneswar, India}
\affil[2]{International Center for Theoretical Sciences (ICTS), Bangalore, India}
\affil[3]{Center for Applicable Mathematics, Tata Institute of Fundamental Research (TIFR-CAM), Bangalore, India}
\affil[4]{Department of Statistics, Lund University, Sweden}

\begin{document}

\maketitle

\begin{abstract}

We propose a representation of the Indian summer monsoon rainfall in terms of a probabilistic model based on a Markov Random Field, consisting of discrete state variables representing low and high rainfall at grid-scale and daily rainfall patterns across space and in time. These discrete states are conditioned on observed daily gridded rainfall data from the period 2000-2007. The model gives us a set of 10 spatial patterns of daily monsoon rainfall over India, which are robust over a range of user-chosen parameters as well as coherent in space and time. Each day in the monsoon season is assigned precisely one of the spatial patterns, that approximates the spatial distribution of rainfall on that day. Such approximations are quite accurate for nearly $95\%$ of the days. Remarkably, these patterns are representative (with similar accuracy) of the monsoon seasons from 1901 to 2000 as well. Finally, we compare the proposed model with alternative approaches to extract spatial patterns of rainfall, using empirical orthogonal functions as well as clustering algorithms such as K-means and spectral clustering.
\end{abstract}

\section{Introduction}

The Indian summer monsoon is a multiscale, multiphysics event that has a profound impact on the food security of a billion people~\cite{gadgilgdp}. Predictions for every year's rainfall from present-day simulations are often wide off the mark~\cite{predfail}, which has serious consequences for the economy of the country. An understanding of the basic dynamical system of the monsoon, including the role of the seasonal migration of the ITCZ and the land-ocean temperature contrast, is an important open problem in climate sciences~\cite{sikkaitcz}. The intra- and inter-annual variability (the temporal variation) as well as the spatial heterogeneity of the monsoon rainfall are quite substantial as are the deviations of local rainfall from the long term average~\cite{gadgilvariability}.

The main aim of the work, presented here and in the companion paper \cite{mitra2018monsoon2}, is to build a data-driven, discrete model of the daily rainfall data at 357 locations over the Indian region, with special emphasis on discovering spatially and temporally small-scale or localized properties and their relation to large-scale or annual, all-India patterns. This is achieved by constructing a Markov random field (MRF) model consisting of two main types of variables (to be described precisely later in section~\ref{ssec:defs}):
\begin{enumerate}
\item ``Hidden'' discrete states (random variables $Z$) at each spatio-temporal location for which the rainfall data is available. The variable $Z$ at each spatio-temporal location, and on each day, takes on one of two values, corresponding to high and low rainfall.
\item Discrete, ``clustering'' variables $U$ for each day and $V$ for each location that respectively encode the spatial and temporal patterns in the discrete variables $Z$ and in the data itself. The variable $U$ for each day takes integer values and essentially indicates the spatial pattern assigned to that day (and similarly, the integer value of $V$ indicates the temporal pattern assigned to that location).
\end{enumerate}
The model we develop is probabilistic, in the sense that the main object of interest is the conditional probability distribution of the random variables $Z,U,V$ conditioned on the rainfall data. We use Gibbs sampling algorithm~\cite{nealmcmc} to sample this probability distribution, and the mode of this distribution is used to study the spatio-temporal patterns for the monsoon rainfall. These patterns are certainly hidden in the data, but are not easy to glean from the data directly. The probabilistic model in terms of the discrete random variables $Z,U,V$ helps us discover these patterns. This paper presents a discussion of the spatial patterns illustrating the utility of the methodology developed here. Further analysis concerning the temporal evolution of the patterns identified in this paper forms one of the central aspects of the companion paper \cite{mitra2018monsoon2}. In particular, we study likely transitions of one spatial pattern to another within a monsoon season. We also analyze the \emph{spatial distribution} of the temporal patterns identified in this paper.

The discussion in this paper is organized as follows. The rest of the introduction summarizes the salient features of the model we study, followed by past data-based approaches to study patterns in spatio-temporal data in earth sciences, ending with a discussion of our main results related to spatial patterns. Then, a complete mathematical description of the MRF model is presented in detail in the following section~\ref{sec:method}. An evaluation of the spatial patterns obtained from this model compared with the patterns obtained from two other commonly used methods, K-means~\cite{kmeans} and spectral clustering~\cite{spect} is in sections~\ref{ssec:eval}-\ref{ssec:patterns}, while a graphical representation of these patterns is given in section~\ref{ssec:picts}. We end with conclusions in section~\ref{sec:conclude}.

\subsection{Main features of the present model}

We construct a Markov random field model with four types of variables $(Z,U,V,X)$ the first three of which are described above, and the fourth one $X$ is a continuous, real-valued random variable denoting the rainfall at each spatio-temporal location. The Markov random field model is shown schematically in figure~\ref{fig:mrfmodel} and described in detail in section~\ref{ssec:mrfmodel}. A novel feature of our model is the introduction of the random variables $U,V$ that denote membership of a specific day or location to a cluster. More explicitly, recall that $U$ takes integer values, say $1, 2, \dots, L$ (for example, $L=10$ later in the paper). Then $U(t) = k$ means that the day number $t$ is assigned the $k$-th spatial pattern out of a total of $L$ spatial patterns. If $t_1, t_2, \dots, t_d$ a total of $d$ number of days which have $U(t_i) = k$, then all these $d$ days essentially form a cluster that have the same spatial pattern of rainfall. Similarly, all locations $s_1, s_2, \dots, s_l$ that are assigned the same temporal pattern $V(s_j) = n$ form a cluster. But note that the assignment of the days and locations to clusters is not deterministic, but  probabilistic.

Thus, in summary, this model has the following two main characteristics:
\begin{enumerate}
\item The MRF for $(Z,U,V,X)$ contains within itself an MRF for the $Z$ variables alone. This $Z$-MRF has edges connecting each location to its geographical neighbours and each day to adjacent days. These edges aid in obtaining a spatio-temporally coherent picture of the rainfall.
\item The full MRF contains edges between the rainfall data $X$ and the clustering variables $U$ as well as edges between the discrete rainfall states $Z$ and the clustering variables. These edges lead the model towards prominent clusters (precisely defined in section~\ref{ssec:patterns}).
\end{enumerate}
The edge potentials for each of these edges as well as priors on the $U,V$ nodes are defined to take into account the qualitative features we expect for the relation between these variables, as explained in detail in sections~\ref{ssec:uv-prior}-\ref{ssec:xzu-edges}.

Once all the edge potentials and priors on some of the variables are defined (section~\ref{ssec:mrfmodel}), the joint distribution for all the variables $p(Z,U,V,X)$ is just a product of all the edge potentials and priors. The main inference step is the sampling of the conditional distribution $p(Z,U,V | X)$, conditioned on the rainfall data $x(s,t)$, which in our case is daily rainfall for 357 locations for eight years 2000-2007. The mode of this conditional distribution is then used as an estimate of hidden state variables $(Z,U,V)$. Each of the clustering variables is naturally related to a discrete as well continuous rainfall pattern, as defined in equations~\eqref{eq:phiu}-\eqref{eq:thetav}, and summarized in table~\ref{tab:acronym}.

In section~\ref{sec:results}, we discuss the properties of the mode of this conditional distribution, including the spatial rainfall patterns of this mode. Note that the conditioning on the observed data leads the model from generic coherent clusters that are present in the prior itself to the coherent clusters that are specific to the rainfall data used for conditioning. We also use these same clusters and test them against a much larger data during 1901-2007 and find that they are robust in a sense described in detail in section~\ref{ssec:patterns}.

Thus in a nutshell, the MRF model edges between $Z$ variables leads to spatio-temporally coherent patterns and the edges to the clustering variables lead to robust clusters that are directly informed by data because of conditioning on the observed rainfall.

\subsection{Relation to past work}

Identification of patterns and clusters is of course a frequently studied problem in the context of climate sciences in general, and for rainfall in particular. Specifically in the context of the Indian monsoon, the problem of understanding the so-called active and break spells (contiguous days of above or below average rainfall) both at the all-India scale and at local scales is an important question. Break spells have been related by~\cite{ramamurthybreak}  to the disappearance of low-pressure zones and eastward winds across India, in the first formal study on break spells. In contrast, \cite{goswamispells} define these phases based on the strength of winds over the Bay of Bengal, while \cite{krishnanbreak} relate these spells to the disappearance of cloud cover over north-west and central India. Active and break spells are defined by \cite{josephspells},\cite{rajeevandataset} and \cite{rajeevanspells} directly by the quantity of rainfall over the Monsoon Zone. Based on the argument that ``intraseasonal variation (of rainfall) is coherent over this zone (the monsoon zone) and the average rainfall over this zone is indeed representative of the rainfall within subregions of the zone'' \cite{rajeevandataset} consider the spatial mean rainfall of the entire zone, whereas \cite{josephspells} divided the zone into eastern and western parts. The mean rainfall for each day is compared against the climatological mean for that date, and accordingly each day is marked as ``active'' or ``break,'' and 3 or more consecutive days marked this way are identified as spells. A similar analysis was done in \cite{krishnamurthyspells}, though they use the all-India spatial mean instead of that over the monsoon zone. Smaller subdivisions of India are studied by \cite{singhspells} to identify and describe regional dry and wet spells. Recently~\cite{windspells} have defined active and break spells with respect to both rainfall and wind over south-western part of peninsula.

The problem of understanding contiguous spatial patterns of rainfall has received much less attention in the context of Indian summer monsoon rainfall. In related contexts, researchers have attempted to quantify spatial coherence in many ways, for example, for African (\cite{spatcoh1}), Indonesian (\cite{spatcoh2}), and only very recently, Indian (\cite{spatcoh3}) rainfall. \cite{paionset} use Principal Component Analysis to select climatic variables as predictors for monsoon onset date, at least 15 days in advance of the actual onset, while~\cite{moumita} uses neural networks to find predictors for annual Indian monsoon rainfall. A commonly used method is based on the empirical orthogonal functions (EOF) of the DRVs, as done for example in~\cite{eofcluster,miso} for other rainfall datasets. Note that, in contrast to the works mentioned above in this paragraph, the method proposed in this paper is specifically targeted to clustering by assigning each day to a spatial pattern and each location to a temporal pattern.

Recent developments in machine learning and data science, along with the availability of high-resolution and accurate climatic data, has resulted in machine learning methods being used to answer questions in climate science in a data-driven way. Modeling extreme events of precipitation has been attempted using Bayesian methods, such as \cite{hierarchicalEV}, \cite{Bayesextr}. Markov Random Fields (\cite{MRF}) are a natural way of handling spatio-temporal data, and they have been used for analysis of ocean surface temperature by~\cite{MRFocean} and detection of large-scale droughts by~\cite{fuMRF}. A recent work that is quite relevant to the current work is~\cite{weathertypes} where six \emph{weather types}, each specified by a spatial pattern of daily low-altitude horizontal winds, are identified using K-means clustering technique over the Pacific.

In the case of the Indian monsoon, not much analysis has been done using machine learning methods. \cite{greeneHMM} makes an important attempt to study the spatio-temporal variations of the Indian monsoon using a Hidden Markov Model (HMM), investigating the spatial patterns of cloud cover and wind directions and their connection with ``active'' and ``break'' phases. One of the main aims of this paper is to fill this gap by providing a robust way of identifying important spatial patterns of the Indian summer monsoon rainfall, using a Markov random field (MRF) model.

\subsection{A summary of our approach and the main results}

We use daily rainfall data of the monsoon seasons (1 June to 30 September) of eight years (2000-2007) over 357 locations across India to obtain a few prominent  spatial patterns. The spatial distribution of each day's rainfall is described by one pattern. These patterns are chosen so as to simultaneously (i) minimize the difference between each day's rainfall distribution (in a discrete representation, as described later) and that of its representative pattern, and (ii) reduce the total number of patterns needed to describe spatial distribution of daily rainfall.

Daily rainfall at a given location is of course a manifestation of an extremely complicated dynamical process involving many physical processes affected by global atmospheric and ocean conditions. This gives rise to the expectation that rainfall over short distances and times will be correlated to each other. To incorporate this in the simplest manner in our model, we use a discrete representation of each day's rainfall at each locations. In the MRF model we impose edge potentials between neighboring spatial locations and times, which has the effect of enhancing spatio-temporal coherence of the discrete variables.

We find that our approach discovers ten spatial patterns, and these patterns are robust over a range of user-chosen parameters. More interestingly, the patterns we obtain from the eight-year data (2000-2007) are representative (to within errors as described below) of over 95 percent of all days in the monsoon seasons of 107 years, from 1901 to 2007. The spatial patterns are coherent, comprehensive and physically realistic.

To demonstrate that our MRF approach is a preferable option to identify spatial patterns in the Indian monsoon rainfall, we obtain for comparison spatial patterns, for the same eight-year rainfall data, by several other approaches, such as by empirical orthogonal functions~\cite{sikkaitcz}, k-means~\cite{kmeans} and spectral means~\cite{spect}. The empirical orthogonal functions are not a direct way of visualising daily rainfall, since the pattern on a given day is obtained as a linear combination of several modes.  

When compared to k-means and spectral clustering, the MRF model gives the least number of prominent patterns which are able to explain the largest number of days in the dataset, and are thus a much better fit to the data compared with the other methods. The MRF patterns are also the most coherent in real space, in the sense that adjacent positions are more likely to have similar values of the discrete rainfall variable. Assignment of each day to one of these patterns by the MRF creates a clustering of the days (indicated by variable $U$). These clusters also cover the largest fraction of the total number of days, compared to clusters of k-means and spectral approaches. The MRF clusters also show the smallest mean Hamming distance of the daily discrete rainfall patterns from that of the cluster each day belongs to, but all the methods show approximately same mean $\ell_2$ distance. Interestingly, when the discrete patterns identified from 2000-2007 data are used for clustering of the data from 1901-2007, the MRF clusters show the smallest mean Hamming as well as $\ell_2$ distances. Since the hamming distance is a natural metric to evaluate discrete patterns, this indicates that our clusters are coherent in the data space.

The patterns we identify, and the sequence of their occurrence, provide an ideal platform to understand many features of the monsoon, and we make a beginning in the companion paper \cite{mitra2018monsoon2}. In that paper we collate the results of this paper, i.e., the spatial and temporal patterns, and thoroughly analyze various aspects of these patterns.


\section{Methodology} \label{sec:method}

In this section, we describe the mathematical model based on Markov random field (MRF) and our main motivation for this model, as well as a description of other methods with which we compare the results from our MRF study. Note that this whole section is devoted to introduce the notation and the model before we come to describing the results in section~\ref{sec:results}.

Consider $S$ locations, which correspond to grid-cells on a rectangular grid system, ordered sequentially first according to longitude and then according to latitude. The indexing scheme has no bearing on the analysis that follows. For each location $s$ with coordinates $(\ell_1,\ell_2)$ we define $\Omega(s)$ as the neighbourhood of this location with coordinates $(\ell_1+a,\ell_2+b)$ where $a\in\{-1,0,1\}$ and $b\in\{-1,0,1\}$ (except for the points on the boundary for which we have to appropriately drop one of $\pm 1$). Consider a sequence of $T$ time-points, each of which corresponds to a day. These days can belong to different years, and the year of day $t$ is indicated by $YY(t)$.

The observed rainfall at location $s$ on day $t$, denoted by $x(s,t)$, is arranged as a matrix of dimension $S \times T$. Corresponding to each day $t$, we have an $S$-dimensional ``daily rainfall vector'' (DRV) $x(t)$ which is a column of that matrix. Similarly each location $s$ is associated with a ``rainfall time series'' (RTS) $x(s)$ which is a row of the matrix.

We want to find a set of ``canonical rainfall patterns'' (CRP) $\{\phi_k\}$ each of which is an $S$-dimensional vector indicating a specific spatial pattern of the rainfall, with the aim of expressing each DRV $x(t)$ in terms of the CRPs. For example, in our model, if $U(t) = u$, then the day $t$ is assigned to the pattern $\phi_u$, whereas in methods such as EOF (described in detail below), $x(t)$ is written as a linear combination of $\{\phi_k\}$. We then wish to identify transition patterns among these CRPs, with the aim of providing a concise view of the progress of monsoon in terms of the temporal evolution of the CRPs. We also want to cluster the days according to the spatial patterns of rainfall and aggregate rainfall, such that some of the clusters belong to ``active days,'' some to ``break days,'' and so on. In order to achieve that task, we want to find a set of ``canonical time series'' (CTS) $\{\theta_l\}$ each of which is a $T$ dimensional vector indicating a specific temporal pattern of the rainfall, with the aim of expressing each RTS $x(s)$ in terms of the CTSs, in analogy of relating DRVs and CRPs. 

We now present a novel approach to the tasks described above. At the heart of this approach is a discrete representation that is easy to interpret and visualize, and allows us to study properties of monsoon rainfall at local and regional scales.


\subsection{Discrete representation: notation and definition} \label{ssec:defs}

Consider a binary latent (random) variable $Z(s,t) \in \{1, 2\}$, which encodes the rainfall amount $x(s,t)$ at location $s$ on day $t$. Its two states $\{1, 2\}$ parameterize two types of distributions for rainfall: $Z(s,t)=1$ corresponds to ``high'' rainfall, and $Z(s,t)=2$ corresponds to ``low'' rainfall at the location $s$ on day $t$. Note that we do not put any \emph{a priori} threshold to pre-define the high and low rainfall amounts. We also note here that the aforementioned distributions are location-specific, to account for the spatial heterogeneity of the rainfall average and variability.

In our model, the rainfall is also considered as a random variable $X(s,t)$, thus interpreting the observed rainfall $x(s,t)$ as a specific realization of this random variable. The precise relation between the continuous $X(x,t)$ and the discrete $Z(s,t)$ is specified later when we describe the Markov random field. For later use and ease of notation, we also denote by $Y(t) = \sum_{s=1}^S X(s,t)$ to be the random variable that is the daily aggregate rainfall on day $t$.

As mentioned earlier, we also wish to find spatio-temporal clusters of (1) the days according to their spatial patterns and (2) the locations according to their temporal patterns. In order to achieve this, we assign to each day $t$ a discrete variable $U(t)$ which takes the value $U(t) = u$ when the day $t$ belongs to cluster number $u$. Similarly, we assign a discrete random variable $V(s)$ to each spatial location $s$, so that its value $V(s) = v$ indicates the membership of location $s$ to a cluster $v$. The total number of either spatial or the temporal clusters is not specified \emph{a priori} but it emerges fron the model. Thus, $u, v \in \mathbb{Z}^+$.

The main objective of introducing these new random variables is to define the CRPs and the CTSs, and their associated discretized versions CDPs and CDSs that were mentioned above. This is done as follows: for each temporal index $u$ (i.e. for each temporal cluster indexed by the integer $u$), the associated spatial patterns CRP and CDP are defined by 
\begin{eqnarray}
\phi_u &=& \textrm{mean}_t\left(X(t) : U(t) = u\right) \,, \nonumber \\
\phi_u^d &=& \textrm{mode}_t\left(Z(t) : U(t) = u\right) \,,
\label{eq:phiu} \end{eqnarray}
where the mean/mode is taken over the days that belong to the cluster $u$. Recall that each $X(t)$ is an $S$-dimensional vector, and hence each $\phi_u$ is also an $S$-dimensional vector. Similarly, for each spatial index $v$ (i.e. for each spatial cluster indexed by the integer $v$), the associated temporal patterns CTS and CDS are defined by
\begin{eqnarray}
\theta_v &=& \textrm{mean}_s\left(X(s) : V(s) = v\right) \,, \nonumber \\
\theta_v^d &=& \textrm{mode}_s\left(Z(s) : V(s) = v\right) \,, 
\label{eq:thetav} \end{eqnarray}
where the mean/mode is taken over the locations that belong to the cluster $v$. Again, $X(s)$ and hence $\theta_v$ are $T$-dimensional vectors.

\begin{table}[t!]
{
\renewcommand{\arraystretch}{1.3}
  \centering
  \begin{tabular}{| p{0.05\columnwidth} | p{0.05\columnwidth} | p{0.08\columnwidth} |  p{0.15\columnwidth} | p{0.32\columnwidth} | p{0.2\columnwidth} |}
    \hline
    Acro\-nym & Sym\-bol & Di\-men\-sion & Num\-ber & In\-ter\-pre\-ta\-tion & Description \\
    \hline
    DRV & $X(t)$       & S=357 & $t \in [1,976]$ & Daily Rainfall Vector & Real-valued; rainfall matrix column\\
    DDV & $Z(t)$       & S=357 & $t \in [1,976]$ & Daily Discretized Vector & Binary form of DRV\\
    CRP & $\phi_u$     & S=357 & $u\in [1,K]$(*) & Canonical Rainfall Pattern & Real-valued; canonical vector to approximate DRVs\\
    CDP & $\phi_u^d$   & S=357 & $u\in [1,K]$ (*) & Canonical Discretized Pattern & Binary form of CRP \\
    RTS & $X(s)$       & T=976 & $s \in [1,S]$ & Rainfall Time Series & Real-valued; rainfall matrix row\\
    DTS & $Z(s)$       & T=976 & $s \in [1,S]$ & Discretized Time Series & Binary equivalent of RTS\\
    CTS & $\theta_v$   & T=976 & $v\in [1,L]$(*) & Canonical Time Series& Real-valued; canonical vector to approximate RTS\\
    CDS & $\theta_v^d$ & T=976 & $v\in [1,L]$(*) & Canonical Discretized Series& Binary form of CTS\\
    \hline
  \end{tabular}
  \caption{The various acronyms and the corresponding mathematical symbols commonly used throughout this and the companion paper. (*) $(K,L)$ denotes the number of spatial and temporal clusters created by the model, and this depends on the data and model parameters mentioned in table~\ref{tab:param}}}
\label{tab:acronym} \end{table}


\subsection{Markov Random Field Model} \label{ssec:mrfmodel}

\begin{figure}[t!]
	\centering
	\includegraphics[width=10cm, height=6cm]{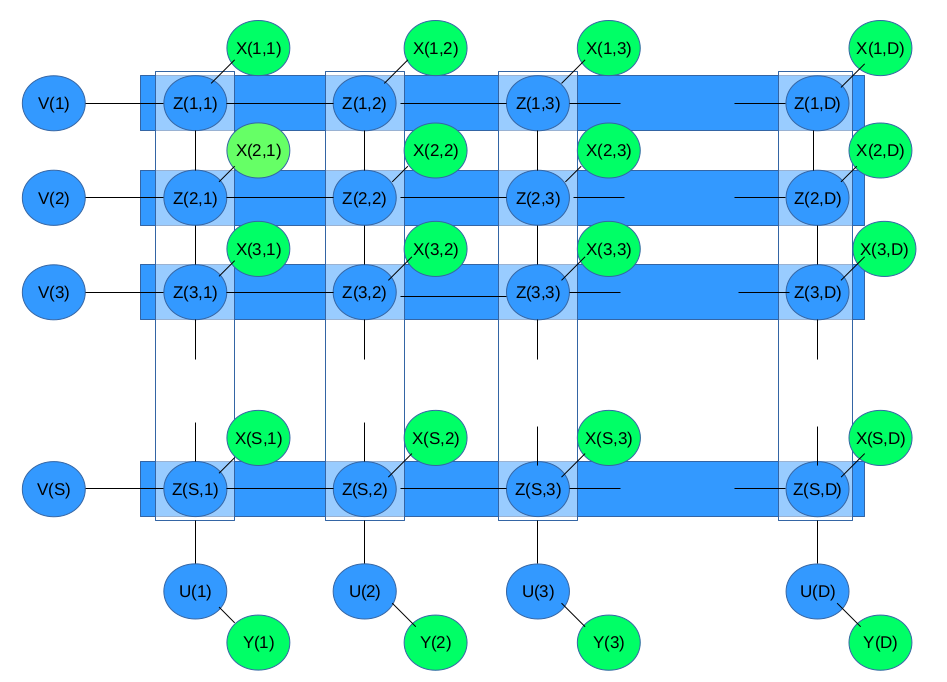}
	\caption{The proposed graphical model for Indian rainfall.  Each column represents one day, each row represents a location. $Z$: binary (discrete) state variable representing high or low rainfall, $X$: rainfall (real valued). Horizontal edges are ``temporal,'' vertical edges ``spatial,''  ``data edges'' connect $Z$ and $X$ nodes. The $U$-nodes represent temporal cluster variables, connected to all local state variables on same day denoted by the vertical rectangles. The $V$-nodes on the left represent spatial cluster variables, connected to all local state variables on same day denoted by the horizontal rectangles.}
\label{fig:mrfmodel} \end{figure}

So far, we have introduced four sets of random variables $(X,Z,U,V)$: the rainfall $X(s,t)$ and a discrete rainfall state $Z(s,t)$ at each location $s$ on day $t$; the cluster membership indicated by $U(t)$ for day $t$ and $V(s)$ for location $s$. The observational rainfall data $x(s,t)$ is a realization of the random variables $X(s,t)$. We aim to make inferences about the discrete variables $Z,U,V$ conditional on the observed data, which can be written as follows.
\begin{equation}
  p(Z,U,V | X) = \frac{p(Z,U,V,X)}{p(X)} \propto p(Z,U,V,X) \,.
\end{equation}
In order to study this conditional distribution, we will use Gibbs sampling method, and hence we will not need the normalization constant $p(X)$ explicitly. Thus the main task now is to describe the joint distribution $p(Z,U,V,X)$ whose definition is in terms of a Markov random field (MRF) with $Z(s,t), U(t), V(s), X(s,t)$ as its nodes. MRF is a graphical model (see~\cite{MRF}) where each random variable is represented by a node, and some pairs of nodes are connected by edges, each of which is associated with an edge potential, and the joint distribution of all the random variables is the product of these edge potentials. In our specific MRF for the $(Z,U,V,X)$ variables, this joint distribution is summarized in the equation below, and the MRF itself is shown schematically in figure~\ref{fig:mrfmodel}. Thus, the joint density $p(Z,U,V,X)$ is summarized in the equation below.
\begin{eqnarray}
  p(Z,U,V,X) &\propto& p_u(U) \times p_v(V) \nonumber \\
&\times& \prod_{s,t}^{S,T}\prod_{t'=t-1}^{t+1}\psi_T(Z(s,t),Z(s,t'))\times \prod_{s,t}^{S,T}\prod_{s'\in \Omega(s)}\psi_S(Z(s,t),Z(s',t)) \nonumber \\
  & \times& \prod_{s,t}^{S,T}\psi_{ST}(Z(s,t),V(s))\times\prod_{s,t}^{S,T}\psi_{SS}(Z(s,t),U(t)) \nonumber \\
&\times& \prod_{s,t}^{S,T}\psi_{DZ}(Z(s,t),X(s,t))\times\prod_{t}^{T}\psi_{DU}(U(t),Y(t))
\label{eq:mrfmodel} \end{eqnarray}
In the above equation, the first two lines contain the prior distributions on the $(Z,U,V)$ nodes (the prior on $Z$ itself being described in terms of the edge potentials for the MRF on $Z$ nodes alone), defined in equations~\eqref{eq:uprior},~\eqref{eq:vprior},~\eqref{eq:coh}. The third line indicates the edge potentials on the $Z-U$ and the $Z-V$ edges, as defined later in equations~\eqref{eq:scale}. The last line indicates the edge potentials on the $X-Z$ and the $X-U$ edges, as defined later in equations~\eqref{eq:zx-edge}-\eqref{eq:ux-edge}. Recall that $Y$ is just a shorthand for the sum of $X$ along the spatial locations, and not a new random variable. We now define in detail each of the terms in the above equation~\eqref{eq:mrfmodel}.

\subsubsection{Prior for the clustering variables $U,V$} \label{ssec:uv-prior}
As stated earlier, the number of clusters is equal to the number of unique values taken by the $U$ and $V$-variables, but these are not decided beforehand. Instead we invoke Chinese Restaurant Process priors on both $U$ and $V$. This is a popular and simple approach to Bayesian nonparametric clustering expressed as a sequential process \cite{pitmanCRP}. The day $t$ can join a cluster from $\{U(1),\dots,U(t-1)\}$ with probability proportional to the number of days assigned to that cluster so far, or form a new cluster with probability proportional to $\gamma$, which is a tuning parameter to be chosen by us. We modify the method slightly, so that the probability of using an existing cluster $k$ is the number of days assigned to it, times the number of years where at least one day was assigned to it. Denote by $n(t,u)=|d: d<t, U(d)=u|$, the number of days assigned to cluster $u$ so far, and $m(t,u)=|y: \exists d: d<t, YY(d)=y, U(d)=u|$, the number of years where at least one day was assigned to cluster $k$. Then the prior distribution of $U$ is given by the following conditional distribution for each $U(t)$-variable:
\begin{eqnarray}
\mathbb{P}\left(\left.U(t)=u\right| U(1),\dots,U(t-1)\right) &\propto& \left\{
\begin{array}{ll} 
 n(t,u)m(t,u) & \textrm{ if } u \in \{U(1),\dots,U(t-1)\} \,, \\
 \gamma & \textrm{ for } u = \max\{U(1),\dots,U(t-1)\} + 1 \,, 
\end{array}
\right. \nonumber \\
\mathbb{P}(U(1) = 1) &=&  1 \,.
\label{eq:uprior} \end{eqnarray}
By modification of Chinese Restaurant Process we ensure that the prominent clusters contain days from different years, i.e. are general across the years rather than forming year-specific clusters. This distribution has the ``rich getting richer" property, i.e. each variable is likely to be assigned to a cluster that is already large. This helps in the identification of a few prominent patterns. We note that this prior distribution implied by the Chinese restaurant process does not take into account any information about the actual rainfall on any given day (i.e. the random variable $X$).

For the spatial cluster variables $V$, we again make use of a similar prior distribution based on Chinese Restaurant Process, where each location $s$ can join a cluster from $\{V(1),\dots,V(s-1)\}$ with probability proportional to the number of locations assigned to that cluster so far (denoted by $n(s,v)=|s': s'<s, V(s')=v|$), or form a new cluster with probability proportional to $\lambda$. Once again, this helps in identification of a few prominent series.
\begin{eqnarray}
\mathbb{P}\left(\left.V(s)=v \right| V(1),\dots,V(s-1)\right) &\propto& \left\{
\begin{array}{ll}
n(s,v) & \textrm{ if } v \in \{V(1),\dots,V(s-1)\} \,, \\ 
\lambda & \textrm{ for } v = \max\{V(1),\dots,V(s-1)\} + 1 \,.
\end{array}
\right. \nonumber \\
\mathbb{P}(V(1) = 1) &=&  1 \,.
\label{eq:vprior} \end{eqnarray}
Let us denote the densities for these prior distributions of $U$ and $V$ as $p_u(U)$ and $p_v(V)$.

\subsubsection{Prior for the discrete rainfall variables $Z(s,t)$} \label{ssec:zprior}
We put a Markov random field prior on the $Z$ variables. In this case, we consider each $Z(s,t)$ variable as a node. For every location $s$ we consider \emph{spatial edges} to the nodes representing its neighboring locations in $\Omega(s)$, i.e. $Z(s,t)$ and $Z(s',t)$ are connected by an edge for each $s'\in \Omega(s)$. Similarly we consider \emph{temporal edges} connecting successive days, i.e. $Z(s,t)$ and $Z(s,t\pm 1)$ are connected by an edge for each location $s$ and each day $t$. The \emph{edge potential functions} on all these edges are described below.

We have already discussed that most climatological phenomena, and especially rainfall, are spatio-temporally coherent. This is particularly true of the discrete (binary) state variable, since it represents the ``state of the climate'' which is a large-scale concept. Hence it is expected to exhibit significantly more spatial and temporal coherence than the actual measured rainfall itself. So, we define the potential functions of temporal and spatial edges between the state variables to promote spatial and temporal coherence. Specifically, we set each function to take high values if the $Z$-variables connected by the edge are equal, and low values if they are different, as given in the equation below.
\begin{eqnarray}
\psi_T(Z(s,t),Z(s,t+1)) &=& f \ \mathds{1}_{\{Z(s,t)=Z(s,t+1)\}} \nonumber \\
\psi_S(Z(s,t),Z(s',t)) &=& g(s,s') \ \mathds{1}_{\{Z(s,t)=Z(s',t)\}} \quad \textrm{for} \quad s' \in \Omega(s) \,.
\label{eq:coh} \end{eqnarray}
Here $\mathds{1}$  is the indicator function and $s'$ is the neighborhood of $s$. The \emph{temporal-coherence parameter} $f$ is assumed constant, while the \emph{spatial coherence parameter} $g$ is specific to the pair of locations $(s,s')$ because not every pair of neighboring locations have the same degree of correlation. For example, locations on the western (windward) side of the Western Ghats mountain range receive very heavy rain during monsoon compared to those on the eastern (leeward) side, even though they are in adjacent grid points, because the mountain range is narrower than the dimensions of our grids. We set $g(s,s')$ to be equal to the correlation coefficient between the RTS at the two locations ($X(s)$ and $X(s')$).

\subsubsection{Edge potentials between discrete rainfall variable $Z$ and clustering variables $U,V$} 
\label{ssec:zuv-edges}

We also construct \emph{spatial scale edges} between the discrete rainfall variables ($Z$-nodes) of each day to the daily cluster variable $U$ for that day, and \emph{temporal scale edges} between the $Z$-nodes of each location to the spatial cluster variable $V$ for that location. The edge potential functions defined on these edges try to align the spatial $Z$-vector of each day (daily discretized vector DDC) to the CDP ($\phi^d$-vector) associated with that day's spatial cluster, and the temporal $Z$-series of each location (discretized time series DTS) to the CDS ($\theta^d$-vector) associated with that day's temporal cluster, as described in equations below.
\begin{eqnarray}
\psi_{SS}(Z(s,t),U(t)) &=& \exp\left(\eta\, \mathds{1}_{\{Z(s,t)=\phi^d(s,U(t))\}}  \right) \,, \nonumber \\
\psi_{ST}(Z(s,t),V(s)) &=& \exp\left(\zeta\, \mathds{1}_{\{Z(s,t)=\theta^d(V(s),t)\}} \right) \,.
\label{eq:scale} \end{eqnarray}

\subsubsection{Edge potentials between rainfall data $X$ and discrete rainfall and clustering variable $Z, U$} \label{ssec:xzu-edges}

We have already mentioned that each $Z$-variable encodes a distribution over the observations $X$. In particular, we choose $X(s,t)\sim \Gamma(\alpha_{sz},\beta_{sz})$ when $Z(s,t)=z$. Thus for each of the two values $z=1,2$, the distribution of $X$ has different means $\alpha_{s1}$ and $\alpha_{s2}$ and also different variance. Thus we naturally include these $X$-variables as nodes into our graphical model by adding \emph{data edges} between each $Z(s,t)$ and $X(s,t)$-node and defining the edge potential of these edges to be the same as the above data distributions, i.e. these Gamma PDFs.
\begin{eqnarray}
\psi_{DZ}(Z(s,t)=z,X(s,t)) &=& \left(X(s,t)\right)^{\alpha_{sz}-1}\exp\left(-\beta_{sz} X(s,t)\right) \,.
\label{eq:zx-edge} \end{eqnarray}
This gives a direct way for the data to inform the $Z$ variables, when we sample the conditional distribution $p(Z,U,V|X)$.

Similarly, we would like to data to directly influence the spatial clustering and hence we include in our MRF the edges between the rainfall variables $X(s,t)$ for a fixed $t$ to the clustering variable $U(t)$ for the same day. In order to describe this edge potential, it is easiest to introduce a variable $Y(t) = \sum_{s=1}^S X(s,t)$. This \emph{aggregate rainfall} is chosen to have a Gaussian PDF with the mean and covariance $(\mu_u,\sigma_u)$ that depends on the cluster $U(t)=u$ to which the day $t$ belongs. Thus each daily cluster is associated with a distribution of the daily aggregate rainfall.
\begin{eqnarray}
\psi_{DU}(U(t),Y(t))     &=& \exp\left(-\frac{1}{2} 
\frac{(Y(t) - \mu_{U(t)})^2}{\sigma_{U(t)}^2}\right) \,.
\label{eq:ux-edge} \end{eqnarray}
This completes our description of the MRF for the $(Z,U,V,X)$ variables and their joint distribution given in equations~\eqref{eq:mrfmodel}-\eqref{eq:ux-edge}. The various parameters that occur in this description are summarized in table~\ref{tab:param}, indicating which of them are set by the user and which are inferred from the model.
\begin{table}[t!]
  \centering
  \begin{tabular}{| p{0.05\columnwidth} | p{0.1\columnwidth} | p{0.1\columnwidth} |  p{0.5\columnwidth}| p{0.1\columnwidth}|}
    \hline
    Sym\-bol & Dimension & Defined in & Description & Set by\\
    \hline
$\gamma$  & 1 & Eq.~\eqref{eq:uprior} & Parameter of Chinese Restaurant Process on $U$ & User\\
$\lambda$ & 1& Eq.~\eqref{eq:vprior} & Parameter of Chinese Restaurant Process on $V$ & User\\
$f$       & 1& Eq.~\eqref{eq:coh} & Temporal Coherence parameter on $Z$ & User\\
$g$       & $S\times\Omega$ & Eq.~\eqref{eq:coh} & Spatial Coherence parameter on $Z$ & User\\
$\eta$    & 1 & Eq.~\eqref{eq:scale} & Relation between $Z$ and $U$ & User\\ 
$\zeta$   & 1 & Eq.~\eqref{eq:scale} & Relation between $Z$ and $V$ & User\\ 
$\alpha$  & 2S& Eq.~\eqref{eq:zx-edge} & Shape parameter of Gamma distribution on $X$ & Model\\
$\beta$   & 2S& Eq.~\eqref{eq:zx-edge} & Scale parameter of Gamma distribution on $X$ & Model\\
$\mu$     & K(*)& Eq.~\eqref{eq:ux-edge} & Mean of Gaussian distribution on $Y$ & Model\\
$\sigma$  & L(*)& Eq.~\eqref{eq:ux-edge} & Standard deviation of Gaussian distribution on $Y$ & User\\
    \hline
  \end{tabular} 
  \caption{The various parameters and the corresponding mathematical symbols related to the proposed model used throughout this and the companion paper. (*) $(K,L)$ is the number of patterns/clusters created by the model.}
\label{tab:param} \end{table}

\subsection{Model Inference by Gibbs Sampling} \label{ssec:gibbs}

Having defined the model with all the random variables and parameters, we now come to the main step: inference of the latent variables $(Z,U,V)$ and estimation of the parameters: $(\alpha,\beta,\mu)$. The other parameters $(f,g,\eta,\zeta,\gamma,\lambda,\sigma)$ are left as user-defined (see table~\ref{tab:param}). The idea of inference is to sample the conditional distribution $p(Z,U,V|X)$ as defined in equation~\eqref{eq:mrfmodel} and using the samples, find the assignment of parameters that maximizes the likelihood. 

We use Gibbs sampling (~\cite{nealmcmc}) which is a Markov Chain Monte Carlo approach. Here, we start with an initial assignment to all the latent variables. Then we visit each random variable one by one, and assign it a value sampled from its conditional distribution, based on all the remaining latent variables. This process is repeated in order to obtain samples from the posterior distributions of the latent variables. We use the mode (obtained from the samples) of this posterior distributions as the optimal estimate of the latent variables. 

In our case, we also have many parameters of these distributions, which are unknown. Estimating these parameters by Expectation-Maximization is computationally very costly. Instead, we use maximum likelihood estimation simultaneously with Gibbs Sampling, where each iteration of latent variable inference is followed by taking a maximum-likelihood estimate of all the parameters based on the current assignment of all the latent variables.

Next, we describe the sampling of each random variable: $Z(s,t)$, $U(t)$ and $V(s)$. In each case, the conditioning set is all the remaining random variables. Here the Markov properties of the Markov random field (see~\cite{MRF}) become important: each variable is independent of all other variables conditioned on its neighbors. This makes the Gibbs sampling step vastly simpler: for sampling each variable we can drop all the terms that do not involve its neighboring nodes.

For sampling $Z(s,t)$, the distribution is as follows:
\begin{eqnarray}
\mathbb{P}\left(Z(s,t)=z|\tilde{Z},U,V,X\right) &\propto& \prod_{t'=t\pm 1}\psi_T(z,Z(s,t')) \times \prod_{s'\in \Omega(s)}\psi_S(z,Z(s',t)) \nonumber \\
&\times& \psi_{ST}(z,V(s)) \times \psi_{SS}(z,U(t)) \times \psi_{DZ}(z,X(s,t)
\end{eqnarray}
where $\tilde{Z}$ denotes all the $Z$ variables except the node $Z(s,t)$ being sampled.

For sampling $U(t)$ and $V(s)$ the distributions are as follows:
\begin{eqnarray}
\mathbb{P}\left(U(t)=u|Z,\tilde{U},V,X\right) &\propto& T'_u(U,t,u)\times\prod_s\psi_{SS}(Z(s,t),u)\times\psi_{DU}(U(t),Y(t)) \nonumber \\
\mathbb{P}(V(s)=v|Z,U,\tilde{V},X) &\propto& T'_v(V,s,v)\times\prod_s\psi_{ST}(Z(s,t),v)
\end{eqnarray}
where $\tilde{U}, \tilde{V}$ denote all the $U,V$ variables except the node being sampled, and $T'_u$ and $T'_v$ are conditional distributions based on Chinese Restaurant Process, which take the same form as $p_u$ and $p_v$ from equations~\eqref{eq:uprior}-\eqref{eq:vprior}, considering value assignments to all the random variables, due to the complete exchangeability property of Chinese Restaurant Process, as explained in~\cite{pitmanCRP}.

Finally, the parameters $(\alpha,\beta,\mu)$ are estimated at each iteration using Maximum-Likelihood, based on the current assignments of the random variables. For example, $\mu_u$ is the sample mean of $Y$ on those days that have been assigned to daily cluster $u$, while estimates of the Gamma parameters $\alpha_{sk},\beta_{sk}$ are obtained using the estimated mean and variance of $X_{s}$ in those days when $Z_s=k$.

As mentioned earlier, the continuous and discrete canonical rainfall patterns $\phi_u$ and $\phi_u^d$ and the continuous and discrete canonical time series $\theta_v$ and $\theta_v^d$ (see table~\ref{tab:acronym}), which are main quantities of interest, are obtained from equations~\eqref{eq:phiu}-\eqref{eq:thetav} with the values of the discrete variables $(Z,U,V)$ obtained from the sample mode and the values of the rainfall are of course obtained from the data.

\subsection{Related Methods} \label{ssec:othermodels}

The tasks of identification of ``canonical'' vectors as well as of patterns and clusters have been studied in various contexts and many different methods have been proposed in data mining literature. We use some of these methods and compare the results obtain from these methods with those obtained from the MRF model described above. 

\paragraph{Using the sample covariance matrix to obtain canonical vectors:}  A commonly used method is based on the empirical orthogonal functions (EOF) of the DRVs, as done for example in~\cite{eofcluster,miso} for other rainfall datasets. This process gives us $S$ vectors, each of dimension $S$, which are the eigenvectors of the sample covariance matrix of the DRVs. We can denote these vectors as $\{\phi^{E}_1,\phi^E_2,\dots,\phi^E_S\}$, which can serve as CRPs. These eigenvectors are indexed in descending order of their associated eigenvalues. 

Each DRV can be expressed as a sum of these eigenvectors and the mean, i.e. $X(t)=\mu+\sum_{j=1}^S\alpha_{tj}\phi^E_j$, where $\alpha$ are regression coefficients and $\mu$ is the mean of $X(t)$. If we want to represent each DRV as a sparse combination of $p$ or less CRPs, we need to solve a problem of sparse linear regression, popularly called LASSO. This can be achieved by solving a regression problem while putting an $\ell_1$-norm regularizer on the regression coefficients. This is formulated as follows:
\begin{equation}
\min_{\alpha}\left( \left\|X_t-\sum_{j=1}^S\alpha_{tj}\phi^E_j \right\|^2+\lambda\sum_{j=1}^S|\alpha_{tj}| \right)
\end{equation}
The parameter $\lambda$ regulates the sparsity of the $\alpha$ vector, i.e. the number of non-zero entries. Increasing $\lambda$ increase the sparsity.

\paragraph{Identification of clusters to obtain canonical vectors:} Another approach to identifying CRPs from the data would be to perform clustering of the data. In particular, we focus on two clustering algorithms: K-means~\cite{kmeans} and spectral clustering~\cite{spect} to partition all the $T$ DRVs $X(t)$ into $K$ clusters, and the $S$ RTSs $X(s)$ into $L$ clusters. From these clusters we compute the cluster mean vectors to serve as CRPs and CTSs.

K-means directly uses the DRVs. Spectral Clustering is done in two ways: once using the exponentials of negative Euclidean distances between DRVs (denoted by Spect1 later), and once using Hamming similarity between thresholded DDVs (denoted by Spect2). Note that the number of clusters to be formed needs to be given as an input to both these methods.

\section{Results} \label{sec:results}

Having defined our model, we now describe our experiments and present our results. The dataset used for this work was published by Indian Institute of Tropical Meteorology. It provides daily rainfall data for the period 1901-2011 at $1^{\circ}-1^{\circ}$ spatial resolution~\cite{rajeevandataset} all over India. The dataset is available on request from \url{http://www.tropmet.res.in/Data%20Archival-51-Page}.

We use daily rainfall data for 8 years, 2000-2007, during the 4 months June-September (122 days each year), over 357 locations across India to identify the CRPs and CDPs. So in our experiments, $S=357$ and $T=122*8=976$. We analyze the daily clusters $U$ and their associated canonical discrete patterns. We also test the identified patterns for an extended period of 111 years (1901-2011) during the 4 monsoon months, over the same 357 locations. Analysis of the spatial clusters $V$ is presented in a companion paper.\cite{mitra2018monsoon2}

\subsection{Evaluation of daily clusters and spatial patterns} \label{ssec:eval}

As already discussed, the number of clusters is not fixed by the user, but learned from the data. However, the user has a control over this number through the parameter $\eta$ (equation~\eqref{eq:scale}). A small value of $\eta$ indicates a large number of clusters, though only a few of them will be prominent (with a significant number of days assigned, and spanning across multiple years). A larger value of $\eta$ will create fewer clusters, and for sufficiently large $\eta$ all days will collapse into a single cluster. In our experiments, we vary this parameter from 5 to 10. The number of clusters obtained for each of these values of $\eta$ is shown in Table~\ref{tab:comp-pc}.

Even for small values of $\eta$, not all the clusters are significant enough, in the sense that the number of days belonging to a cluster may be too small. Hence, we define \emph{prominent clusters} to be those having members from at least 5 of the 8 years. We see that even with different $\eta$ and hence different number of total clusters, the number of prominent clusters is approximately constant at around 10. For comparison of these results with those from K-means~\cite{kmeans} and spectral clustering~\cite{spect} (which require specification of number of clusters), we specify the same number of clusters as obtained for that specific value of $\eta$ and then find the number of prominent clusters using the same definition above. The number of prominent clusters for each of these methods (our model, K-means, and two spectral clustering algorithms denoted by SP1 and SP2, as mentioned in section~\ref{ssec:othermodels}) is reported under the columns titled ``\#PC'' in Table~\ref{tab:comp-pc}.

The number of days (out of a total of $T=976$) covered by the prominent clusters identified by each method indicates the significance of these clusters. This number is shown under the columns titled ``PC coverage'' in Table~\ref{tab:comp-pc}.

An important criteria to evaluate clustering is the intra-cluster homogeneity. In this case, it is expected that each cluster should be uniform with respect to the daily aggregate rainfall denoted by $Y(t)$. A cluster's uniformity with respect to $Y$ can be measured by standard deviation of the $Y$-values assigned to it, and it is also reported in Table~\ref{tab:comp-pc} under the columns titled ``std(Y).''

From Table~\ref{tab:comp-pc}, it is clear that the proposed model forms the least number of prominent clusters compared to the other clustering methods, and still covers a fairly large (though not necessarily the largest) number of daily vectors in them. On the other hand, the average number of days per prominent clusters is highest in case of the proposed model. At the same time, this model is able to maintain homogeneity of the clusters compared to the other methods, which is indicated by the smallest standard deviation for the aggregate rainfall $Y$ for days that belong to each cluster.
	\begin{table}
		\centering
		\begin{tabular}{| c | c | c | c | c | p{0.06\textwidth} | p{0.06\textwidth} | p{0.06\textwidth} | p{0.06\textwidth} | c | c | c |}
			\hline
			$\eta$ & \multicolumn{4}{| c |}{\#PC} & \multicolumn{4}{| c |}{PC coverage (average)}& \multicolumn{3}{| c |}{std(Y)}\\
			\hline
			(\#clusters) & MRF & KM & SP1 & SP2 & MRF & KM & SP1 & SP2 & MRF & KM & SP1\\
			\hline
			5 (146) & \textbf{11} & 26 & 38 & 30 & \textbf{556} (\textbf{50.5}) & 516 (19.8) & 514 (13.5) & 378 (12.6) & \textbf{1.06} & 1.28 & 1.5\\
			7 (65)  & \textbf{11} & 26 & 43 & 44 & 786 (\textbf{71.5}) & 735 (28.3) & \textbf{816} (19.0) & 800 (18.2) & \textbf{1.07} & 1.73 & 1.77\\
			8 (36)  & \textbf{10} & 20 & 34 & 34 & 866 (\textbf{86.6}) & 862 (43.1) & \textbf{953} (28.0) & 928 (27.3) & \textbf{1.06} & 1.86 & 1.76\\
			9 (24)  & \textbf{10} & 18 & 22 & 23 & 938 (\textbf{93.8}) & 951 (52.8) & \textbf{953} (43.3) & 944 (41.0) & \textbf{1.05} & 2.08 & 1.85\\
			10 (15) & \textbf{11} & 16 & 15 & 15 & 966 (\textbf{87.8}) & 965 (60.3) & \textbf{976} (65.1) & \textbf{976} (65.1) & \textbf{1.22} & 2.27 & 1.87\\
			\hline 
		\end{tabular}
		\caption{Comparison of daily cluster properties, by varying the number of clusters through $\eta$ parameter of the proposed model. \#PC denotes number of prominent clusters (spanning at least 5 years), and PC coverage denotes number of days (out of 976) assigned to the prominent clusters, and the number in parenthesis gives the average number of days per prominent cluster. The last columns give the standard deviation of the aggregate daily rainfall for days assigned to a cluster. The best performing value is highlighted in bold.}
\label{tab:comp-pc} \end{table}

The clustering of the days by any method can be used to identify real-valued CRPs which are analogous to the CDPs. The CRP corresponding to each daily cluster is the mean of the DRVs of the days assigned to that cluster. Similarly, for each CRP (identified by K-Means or Spec1) we can have an analogous CDP by thresholding against location-wise mean daily rainfall.

The effectiveness of the canonical patterns -- CDPs $\phi_u$ and CRPs $\phi_u^d$ -- can be measured by comparing each day's rainfall vector (DRV) and discrete vector (DDV) to the canonical pattern associated with that day's corresponding cluster. To compare real-valued $\phi$ and $X$ vectors, we use $\ell_2$ distance, i.e.,
\begin{equation}
\ell_2(\phi)= \sum_t||X(t)-\phi(U(t))||_2 \,.
\label{eq:ell2} \end{equation}
Similarly for binary-valued $\phi^d$ and $Z$ vectors, we use Hamming distance (number of elements of a vector that are different from those of another vector) with $Z$, i.e.,
\begin{equation}
\textrm{Hamm}(\phi_d)= \sum_{s,t}I(Z(s,t)\neq\phi_d(U(t))) \,.
\label{eq:hamm} \end{equation}
Once again, we report these in Table~\ref{tab:comp-dist}. For $\ell_2$-distance between DRV and CRP we compare the proposed model, K-means and Spect1,  while for Hamming distance between DDV and CDP we compare the proposed model, K-means and Spect2.

Yet another measure that we use to compare the DRVs to the CRPs is the error in the daily aggregate rainfall, i.e. $Y$. Recall that each $\phi_u$ is an $S$-dimensional vector. Hence we can associate each canonical pattern (CRP) with an aggregate rainfall volume $\hat{\phi_u} = \sum_s\phi_{u}(s)$. The corresponding error measure is
\begin{equation}
\textrm{Agg}(\phi)=\sum_t|Y_t-\hat{\phi}_{U(t)}| \,.
\label{eq:agg} \end{equation}
Once again, this can be computed for CRPs identified by the proposed method, K-means and Spec1 and is shown in table~\ref{tab:comp-dist}.

With respect to the binary patterns $\phi^d$, table~\ref{tab:comp-dist} shows that those produced by the proposed method fit the DDVs better than K-means and spectral clustering (Spect2), even though the latter explicitly defines pairwise similarity measures using Hamming Distance. With respect to the $\ell_2$ distance of DRVs from cluster centers (CRP), K-means and spectral clustering (Spect1) understandably do better than the proposed model as they explicitly aim to minimize this quantity by using $\ell_2$ distances for objective function/pairwise similarity measure. The proposed model does not use it explicitly, but does not lag far behind. In terms of the aggregate rainfall, the clusters defined by the proposed model perform better than the K-means or Spec1.

The results in Table~\ref{tab:comp-dist} can also be interpreted using Akaike Information Criteria (AIC) for model comparison. For clustering problems, the AIC value of a model is given by $AIC=2K-2log(P)$ where $K$ is the number of clusters and $P$ is the model likelihood, and a model with a smaller value of AIC is better. If we consider a Gaussian model likelihood, i.e. $X(t)\sim\mathcal{N}(\phi_{U(t)},I)$, then the log-likelihood is equal to the mean $\ell_2$-distance between each DRV $X(t)$ and the corresponding CRP $\phi(U(t))$. This is the likelihood associated with K-means, and undoubtedly it performs best. But if we consider the Hamming model likelihood, i.e. $\prod_{s}\exp\left(\eta\mathds{1}_{Z(s,t)=\phi^d(s,U(t))}\right)$ as used in the proposed model (see equation~\ref{eq:scale}, then the log-likelihood is equal to the mean Hamming distance between each DDV $Z(t)$ and the corresponding CDP $\phi^d(U(t))$. In this case the proposed model predictably gives the best results.
	\begin{table}[t!]
		\begin{tabular}{| c | c | c | c | c | c | c | c | c | c |}
			\hline
			$gg$  & \multicolumn{3}{| c |}{$\ell_2(\phi)$} & \multicolumn{3}{| c |}{Hamm($\phi_d$)} &  \multicolumn{3}{| c |}{Agg($\phi$)}\\
			\hline
			(\#clusters) & MRF & KMeans & Spect1 & MRF & KMeans & Spect2 & MRF & KMeans & Spect1\\
			\hline
			5 (146) & 215 & \textbf{184} & 201 & \textbf{31} & 58 & 57  & \textbf{0.92} & 1.05 & 2.47\\
			7 (65)	& 239 & \textbf{211} & 222 & \textbf{46} & 65 & 65  & \textbf{0.92} & 1.3  & 2.6\\
			8 (36)  & 234 & \textbf{225} & 234 & \textbf{53} & 69 & 70  & \textbf{0.96} & 1.47 & 2.63\\
			9 (24)  & 251 & \textbf{232} & 240 & \textbf{58} & 71 & 72  & \textbf{0.86} & 1.62 & 2.61\\
			10(15)  & 255 & \textbf{240} & 247 & \textbf{60} & 73 & 75  & \textbf{0.86} & 1.63 & 2.67\\
			\hline 
		\end{tabular}	
		\caption{Comparison of daily cluster properties, by varying the number of clusters through $\eta$ parameter of the proposed model. $\ell_2(\phi)$ is the mean $\ell_2$-distance of DRVs to CRP $\phi$ of corresponding cluster (equation~\eqref{eq:ell2} and $\textrm{Hamm}(\phi_d)$ is the mean Hamming distance of DDVs to CDP $(\phi_d)$ of corresponding cluster (equation~\eqref{eq:hamm}). $\textrm{Agg}(\phi)$ is the mean absolute error of the daily aggregate rainfall $Y$ (equation~\eqref{eq:agg}).}
\label{tab:comp-dist} \end{table}

\subsection{Prominent patterns} \label{ssec:patterns}

We have already introduced the term \emph{prominent cluster}. The corresponding spatial patterns (CRP and CDP) will be called \emph{prominent pattern}, one which appears on at least one day in at least 5 of the 8 years considered. Table~\ref{tab:comp-pc} shows that the proposed method always produces about 10 such prominent patterns for a wide range of the parameter $\eta$, and we find that these patterns are constant across different values of $\eta$. These patterns also cover a significant number of days in this period, which is as high as $95\%$ for $\eta=9,10$. This indicates that these 10 patterns are quite robust and frequent, and there are only a small number of days per year which do not conform to these patterns. We also find that these days are the ones having excessive rainfall.

In the remainder of the paper, we will evaluate only these prominent patterns. Since K-Means and Spec1 produce much greater number of prominent patterns compared to the 10 produced by the proposed method (see table \ref{tab:comp-pc}), for the sake of fair comparison we run these algorithms with the number of clusters chosen such that the number of prominent patterns identified by them is also around 10.

The spatial patterns we have extracted are based on only 8 years of data from 2000-2007. The question arises: are these patterns general enough? Can these be used to approximate DRVs from years beyond this period? Accordingly, we considered the $\ell_2(\phi)$ and $\textrm{Hamm}(\phi_d)$ measures across the period 1901-2011, i.e., the sum over time in equations~\eqref{eq:ell2}-\eqref{eq:hamm} is taken from 1901-2011. 

For this purpose, we run the model again using the DRVs from this period, but we use the same $\theta$ and $\phi$ which were estimated earlier from the period 2000-2007. During the inference process based on Gibbs Sampling we do not update these parameters. In effect, we estimate DDVs corresponding to the DRVs of this period, and then try to approximate them with the same set of CDPs as already discovered. In case of the other approaches like K-means and Spec1, we try to approximate these DRVs with the CRPs computed from the period 2000-2007.
The results are shown in Table~\ref{tab:fullfit}, and we find the prominent CRPs from the proposed model fit slightly better than those identified by the other two methods.
	\begin{table}[t!]
		\centering		
		\begin{tabular}{| c | c | c || c | c | c || c | c | c |}
			\hline
			\multicolumn{3}{| c ||}{$\ell_2(\phi)$} & \multicolumn{3}{| c ||}{Hamm($\phi_d$)} &  \multicolumn{3}{| c |}{Agg($\phi$)}\\
			\hline
			MRF & KMeans & Spect1 & MRF & KMeans & Spect2 & MRF & KMeans & Spect1\\
			\hline
			\textbf{261} & 263 & 262 & \textbf{104} & 202 & 187 & \textbf{0.49} & 0.7 & 0.75  \\
			\hline 
		\end{tabular}
		\caption{Measures of how well the spatial patterns (CRP and CDP) computed over the period 2000-2007 can approximate the daily vectors (DRVs and DDVs) across the period 1901-2011. Three measures are considered: $\ell_2(\phi)$, $\textrm{Hamm}(\phi_d)$, and $\textrm{Agg}(\phi)$ are define in equations~\eqref{eq:ell2}-\eqref{eq:agg}}
	\label{tab:fullfit} \end{table}

Finally, in Table~\ref{tab:coher} we study the \emph{spatial coherence} of the CDPs and CRPs identified by different measures. This is done separately for each pattern by comparing the value at each location to those of the adjacent locations. In case of CRPs this is measured as $\sum_{k,s}\sum_{s'\in \Omega(s)}{\frac{|\phi_{s'k}-\phi_{sk}|}{|\phi_{sk}|}}$, while in case of CDPs it is measured as $\sum_{k,s}\sum_{s'\in \Omega(s)}I(\phi^d_{s'k}\neq\phi^d_{sk})$. We find that the most spatially coherent CDPs are produced by the proposed method, compared to K-means, Spec2 and also Empirical Orthogonal Functions (EOF). Spec1 produces the most spatially coherent CRPs.
	\begin{table}[t!]
		\centering		
		\begin{tabular}{| c | c | c | c |}
			\hline
			\multicolumn{4}{| c |}{$spch(\phi_d$)}\\
			\hline
			MRF & KMeans & Spect2 & EOF\\
			\hline
			\textbf{0.07} & 0.16 & 0.14 & 0.13\\
			\hline 
		\end{tabular}
		\caption{Measure of spatial coherence of the CDPs discovered by the different methods.}
	\label{tab:coher} \end{table}

\subsection{Graphical Representation of spatial patterns} \label{ssec:picts}

\begin{figure}[t!]
	\centering
	\includegraphics[width=\textwidth]{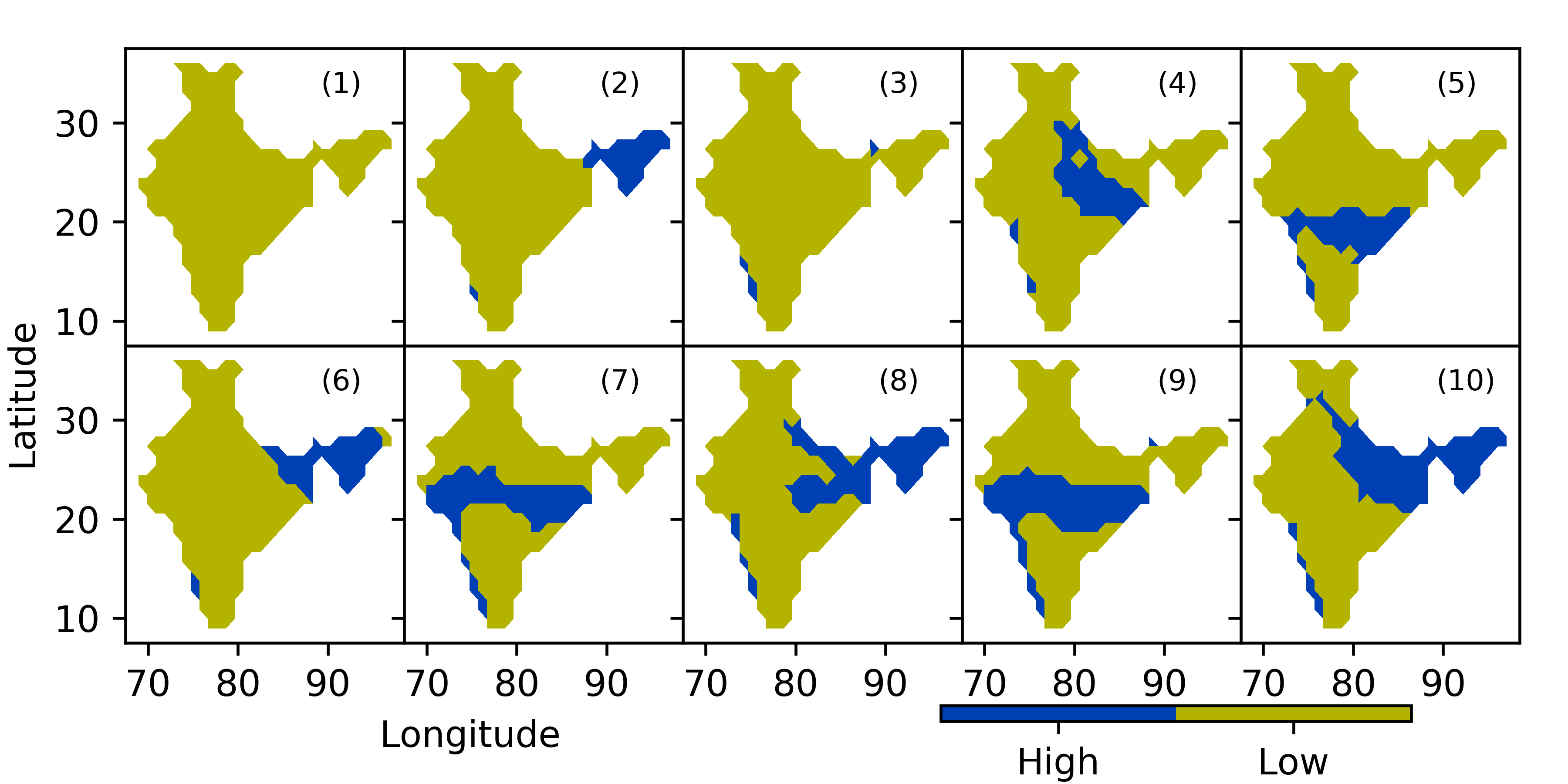}
	\includegraphics[width=\textwidth]{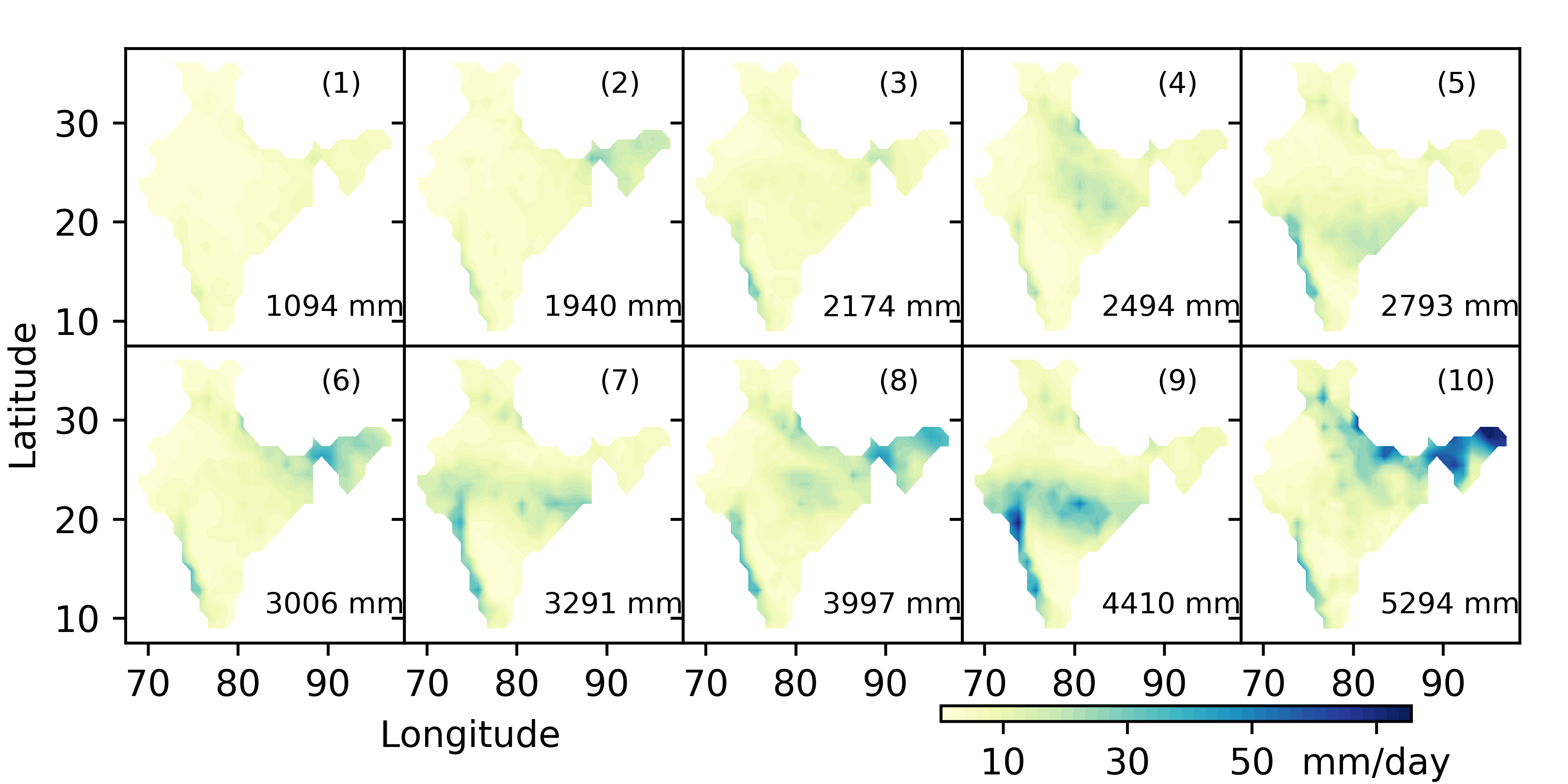}
	\caption{Prominent canonical discrete patterns (CDP) and the corresponding canonical rainfall patterns (CRP) identified by the proposed model. The numbers on bottom right show the total rainfall in mm/day for that pattern.}
    \label{fig:mrf-cdp-crp} 
\end{figure}

As already discussed, each cluster produced by the proposed model and by spectral clustering with Hamming distance (Spec2) is associated with a CDP, and each cluster produced by the proposed model, by k-means, and by spectral clustering with Euclidean distance (Spec1) is associated with a CRP. In all the settings shown above, the proposed model produces about 10 prominent clusters. Corresponding to each of these, we identify 10 prominent CDPs $\phi_u^d$ and CRPs $\phi_u$, which are, respectively, binary and real-valued vectors of dimension $S$ corresponding to the $S$ locations. Thus these vectors can be shown on a map, as done in Figure~\ref{fig:mrf-cdp-crp}. Each panel corresponds to one CDP (top figure) or CRP (bottom figure), where the green locations are in state 2 (low rainfall) while blue ones are in state 1 (high rainfall).

The spatial coherence of these wet and dry zones are very notable. Clearly, some patterns are associated with pre-onset period or break spells where there is no or very little rainfall, while some others are associated with active spells with the central region (``monsoon zone'') turning active. Some of the patterns look similar in the discrete representation, for example the seventh and ninth CDPs, but the aggregate rainfall volume associated with them are different. Note that some locations in the north-west and south-east are not active in any of these prominent patterns. These are the regions that tend to remain dry during the monsoon. Their rare rainy days are covered by the non-prominent patterns.

\begin{figure}[t!]
	\centering
	\includegraphics[width=\textwidth]{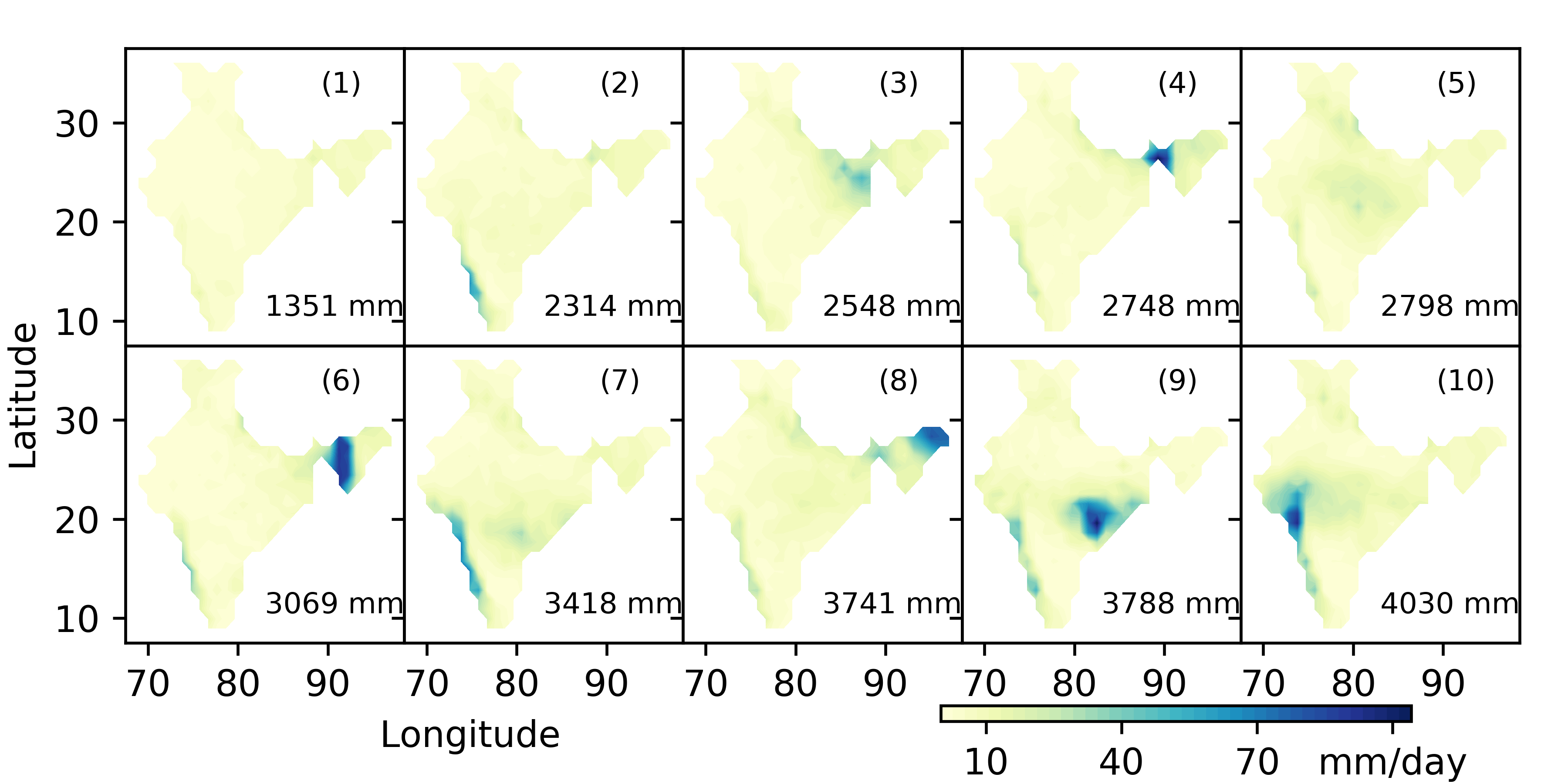}
	\caption{Canonical Rainfall Patterns (CRP) corresponding to 10 prominent clusters found by K-means}
\label{fig:kmean-crp} \end{figure}

\begin{figure}[t!]
	\centering
	\includegraphics[width=\textwidth]{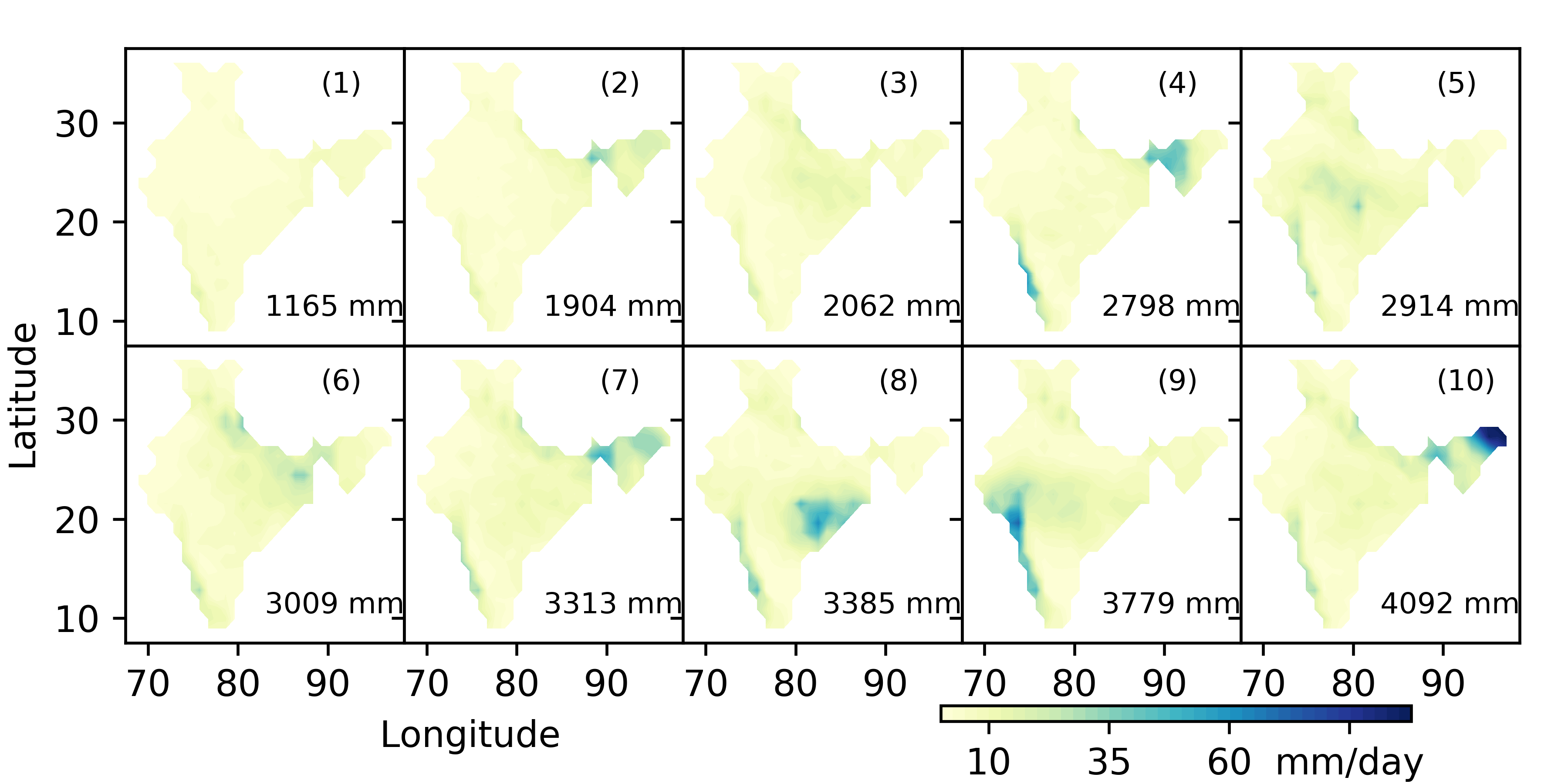}
	\caption{Canonical Rainfall Patterns (CRP) corresponding to 10 prominent clusters found by Spectral Clustering (Spec1)}
\label{fig:spec1-crp} \end{figure}

\begin{figure}[t!]
	\centering
	\includegraphics[width=\textwidth]{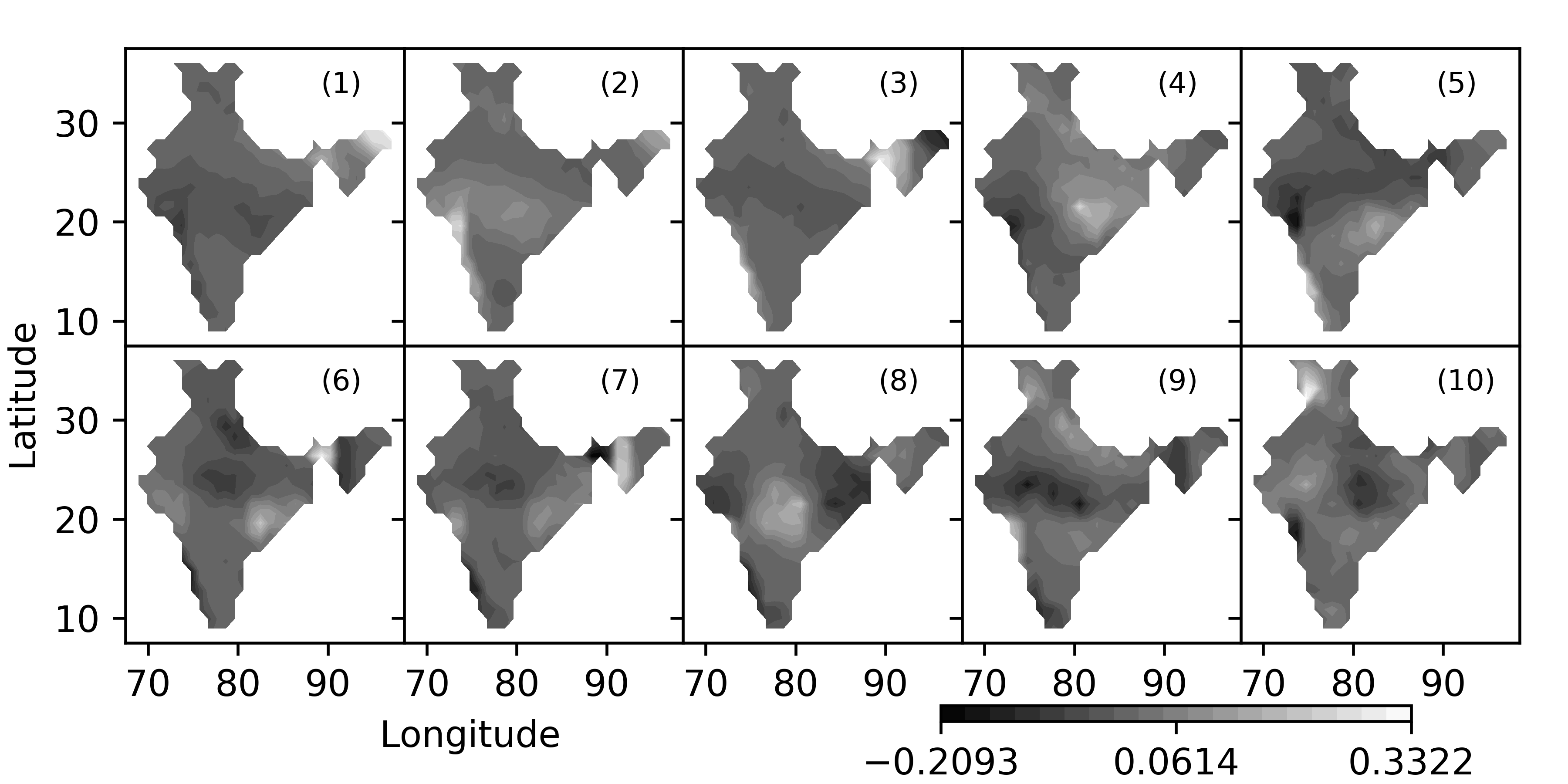}
	\caption{The first 10 Empirical Orthogonal Functions (EOF)}
\label{fig:eof-crp} \end{figure}

In order to compare the CRPs obtained from our mode, we also show the prominent CRPs identified by K-means and Spec1, as well as the first 10 empirical orthogonal functions (EOF), in Figures~\ref{fig:kmean-crp}, \ref{fig:spec1-crp}, and \ref{fig:eof-crp} respectively.

\begin{figure}[t!]
	\centering
	\includegraphics[width=0.3\textwidth]{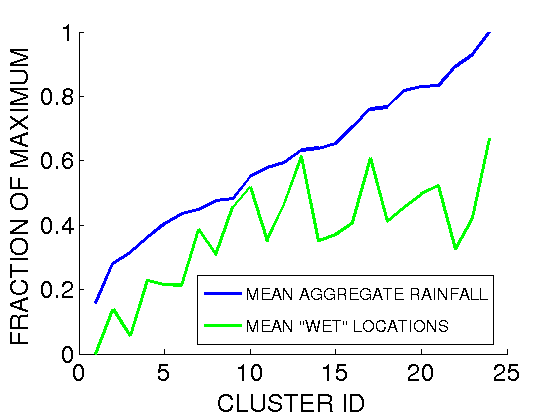}
    \includegraphics[width=0.3\textwidth]{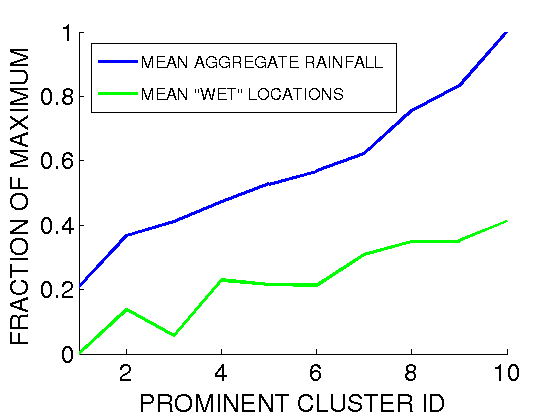}
    \includegraphics[width=0.3\textwidth]{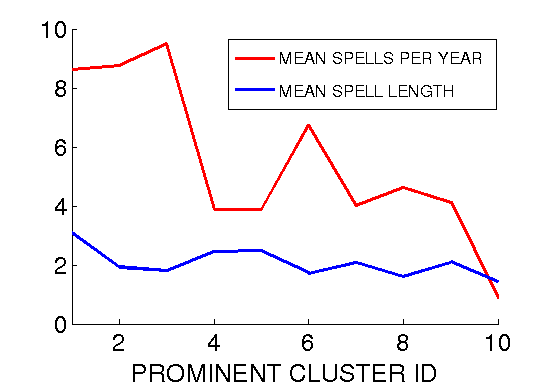}
	\caption{Properties of the clusters/patterns identified by the proposed model. Left: mean daily aggregate rainfall for the days assigned to each cluster, and mean fraction of locations that are ``wet'' on such days. Middle: same analysis for the 10 prominent patterns mentioned above. Right: Mean number of ``spells'' of each pattern per year, and mean length of such spells}
\label{fig:agg} \end{figure}

We also study some properties of these clusters and the associated spatial patterns. Each of them corresponds to a distribution over daily aggregate rainfall $Y$, and in figure~\ref{fig:agg} we plot the mean of this distribution for each of the clusters, as $\mu_k=\text{mean}\{Y_t: U(t)=k\}$. This is done for all the 24 clusters obtained by setting $\eta=9$ (see Table~\ref{tab:comp-pc}), and also the 10 prominent clusters among them. Also, each CDP has a fraction of the locations in ``wet'' state $\phi_{u}(s) = 1$, and in figure~\ref{fig:agg} we also plot this fraction. The plots show clearly the variation of both these quantities across the patterns, which indicates that some patterns are associated with ``active spells'' and some with ``break spells.'' 

From the assignments of $U$ we see that each CDP tends to persist for a few days, like a spell. Across the entire period, we identify such spells for each CDP, and compute the mean spell length, i.e. mean number of days for which each CDP persists. The number of such spells, and the mean spell length for each CDP are plotted in the rightmost panel of Figure~\ref{fig:agg}. It shows us that there are more spells of dry patterns per year than spells of wet patterns, but the mean spell length is more or less uniform for all patterns.

\section{Conclusion} \label{sec:conclude}

In this paper, we proposed a model that is capable of identifying the common spatial patterns of daily rainfall over India during monsoon. The model is based on advanced Bayesian nonparametric methods, which has not received significant attention in the field of geo-sciences. The Markov random field (MRF) model we propose allows us to identify patterns based on the data, while incorporating domain knowledge related to spatial and temporal coherence. The model creates a discrete representation of daily rainfall, which is easy to visualize and interpret. 

The rainfall distribution on each day is assigned to one of these patterns, and thence the days may be grouped into clusters. We compared our method to an alternative approach used earlier: namely identification of empirical orthogonal functions (EOF). We also performed analysis using clustering algorithms like K-means and Spectral Clustering, and showed that our MRF based approach produces clusters that are more homogeneous and coherent both in the data space and the geographical space. The spatial patterns we find are more representative of the daily spatial distributions of rainfall providing a better fit to data, and are quite robust. The patterns we identified are interpretable to climate scientists by virtue of their spatial coherence. Additionally, in our approach the data vector of each day is approximated by a single pattern unlike EOF which produces a linear combination of patterns.

We thus show that only 10 spatial patterns can represent nearly $95\%$ days of each monsoon season with reasonable accuracy. These patterns were identified from only 8 years (2000-2007) but they fit well on daily data from over a hundred years. In a companion paper,\cite{mitra2018monsoon2} we will discuss the temporal characteristics of these patterns, showing that some of them to be more frequent in the early and late monsoon months (June, September) while the others are more frequent in the peak monsoon months (July, August), and also study about homogeneous zones on the landmass which are also identified by this model.

When identifying prominent patterns (i.e. those that occur in at least 5 out of 8 years, see section~\ref{ssec:patterns}) using these three methods, the MRF model gives the least number of such prominent patterns which on an average are able to explain the largest number of days in the dataset - the CRPs obtained by the model are much better fit to the data compared with the other methods. The number and qualitative features of the patterns is also almost constant across a range of values of the parameters that are part of the model - these CRPs are quite robust. These comparisons are discussed in detail in section~\ref{ssec:eval}. The MRF clusters are also the most coherent in real space, in the sense that adjacent positions are likely to have similar values of the discrete rainfall variable (see table~\ref{tab:coher}).

The MRF model naturally provides a clustering of the days. These clusters cover the largest fraction of the total number of days (see table~\ref{tab:comp-pc}). The clusters we find also show the smallest mean Hamming distance of the individual cluster members to the CDPs compared with other methods, but all the methods show approximately same mean $\ell_2$ distance. Interestingly, when the CDPs identified from 2000-2007 data are used for clustering of the data from 1901-2007, the MRF clusters show the smallest mean Hamming as well as $\ell_2$ distances - the clusters we find are quite coherent in the data space, certainly when using Hamming distance which is the natural metric to use for discrete patterns.

\section*{Acknowledgements}
AM was with ICTS-TIFR, Bangalore, India when most of this work was done. AM, AA, SV would like to acknowledge support of the Airbus Group Corporate Foundation Chair in Mathematics of Complex Systems established in ICTS-TIFR and TIFR-CAM. AM, AA would like to thank The Statistical and Applied Mathematical Sciences Institute (SAMSI), Durham, NC, USA where a part of the work was completed.

\bibliography{discrete-monsoon}
\bibliographystyle{plain}


\end{document}